\documentclass[aps,prb,reprint,floatfix,amsmath,amssymb,amsfonts,superscriptaddress]{revtex4-2}

\usepackage{hyperref}
\hypersetup{colorlinks=true,allcolors=blue}
\usepackage[all]{hypcap}
\usepackage{orcidlink}
\usepackage{graphicx}

\begin{document}
\author{Pedro Portugal\,\orcidlink{0009-0009-5151-4400}}
\affiliation{Department of Applied Physics, Aalto University, Finland}

\author{Riku Tuovinen\,\orcidlink{0000-0002-7661-1807}}
\affiliation{Department of Physics, Nanoscience Center, University of Jyväskylä, Finland}

\author{Christian Flindt\,\orcidlink{0000-0002-7223-8400}}
\affiliation{Department of Applied Physics, Aalto University, Finland}

\title{Generation of heat pulses in mesoscopic conductors using light fields}
\date{\today}
\begin{abstract}
We propose to generate heat pulses in mesoscopic conductors using light fields. In contrast to single-electron excitations such as levitons, which are created by accurate voltage drives, our approach relies on modulating the temperature of an electronic reservoir. To this end, we show that the interactions with a light field can induce a controllable time-dependent temperature in an electrode. The temperature modulations generate charge-neutral heat pulses that can be emitted into a mesoscopic conductor and detected in the outputs. We illustrate our approach by evaluating the time-dependent currents and their fluctuations using a tight-binding model of two electronic reservoirs connected by a quantum point contact. Our work establishes a route towards on-demand caloritronics, where energy rather than charge carries quantum information, and it paves the way for probing time-resolved heat transport and quantum coherence in thermally driven conductors.
\end{abstract}

\maketitle

\section{Introduction}

Electron quantum optics has in recent years emerged as a novel framework for investigating quantum coherence and correlations of fermionic excitations in mesoscopic conductors~\cite{Bocquillon:2014}. Inspired by quantum optics with photons, electron quantum optics explores the emission, propagation, and interference of few-electron wavepackets in solid-state systems. Key experiments include the realization of on-demand single-electron sources in quantum Hall edge channels~\cite{Feve:2007}, the observation of minimal excitations known as levitons generated by lorentzian voltage pulses~\cite{Dubois2013a,Jullien:2014}, recently also in graphene~\cite{Assouline2023,chakraborti2025}, and the demonstration of Hong–Ou–Mandel interference and fermionic antibunching~\cite{Bocquillon:2013,ouacel2025}. In combination, these experiments show that time-dependent electric currents can be decomposed into discrete, coherent few-particle excitations~\cite{roussel2021}, opening a direct route to the manipulation of quantum information encoded in flying qubits~\cite{edlbauer2022,pomaranski2024}.

While experiments in electron quantum optics have focused on time-dependent voltages, investigations of heat transport in coherent conductors have mainly involved measurements with constant temperature differences~\cite{Pekola:2021}. Groundbreaking work has observed the quantization of the heat conductance of the individual transmission channels in a mesoscopic conductor~\cite{jezouin2013}, and subsequent heat measurements have been extended to topological systems, which may exhibit half-integers of the thermal conductance quantum~\cite{Banerjee:2018}. These experiments have established the precise control and detection of heat currents in the steady-state regime and clarified their relevance for quantum-limited thermometry and emerging quantum technologies~\cite{Pekola2013,Majidi2024}. However, setups with fixed temperatures only probe the stationary heat flow, while the microscopic dynamics of heat exchange, in particular, how heat is emitted, carried, and interferes on fast timescales, remain a largely unexplored territory.

\begin{figure}
	\centering
\includegraphics[width=0.95\columnwidth]{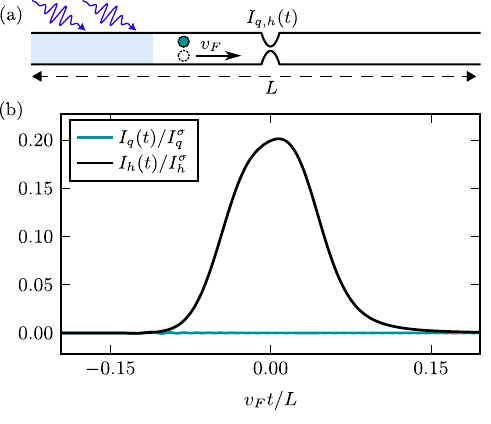}
	\caption{Generation of heat pulses in a mesoscopic conductor. (a) A high-frequency light field generates a heat pulse, which travels at the Fermi velocity~$v_F$ along a one-dimensional conductor of length $L$. The pulse is charge-neutral and only carries heat. (b) The pulse produces a time-dependent heat current, $I_h(t)$, at the quantum point contact but no electric current, $I_q(t)$. Details of the calculation are described in Sec.~\ref{sec:heatpulses}.}
	\label{fig:fig1}
\end{figure}

Several theoretical works have investigated heat transport in coherent conductors, also with time-dependent temperatures \cite{eich2014,tatara2015,biele2015,eich2016,lozej2018,covito2018,karaslimane2020,portugal2024,lopez2024,acciai2025}. Heat currents can be generated by coupling a driving field to the total energy of an electrode, measured from the Fermi level, similarly to how an electric current can be generated by a driving field that couples to the total charge in the electrode. In the context of heat currents, the driving field is sometimes referred to as a Luttinger field, following the early work by Luttinger on thermal transport with static temperature gradients~\cite{luttinger1964}. However, unlike electric currents, where the applied voltage is the field that couples to the total charge, it is less clear how a Luttinger field can be implemented in practice. The purpose of the present work is therefore to propose a way to realize a dynamic Luttinger field and thereby generate time-dependent heat currents.

The basic idea of our proposal is illustrated in Fig.~\ref{fig:fig1}(a), showing two electrodes connected by a quantum point contact. We shine light on one of the electrodes to generate heat pulses. Specifically, by using a frequency that is larger than the bandwidth of the host material, it is possible to dynamically modify the Fermi velocity and thereby change the electronic temperature. This is not an incoherent process involving dissipative heating or cooling of the electrode. Rather, it is a coherent process, which resembles the adiabatic compression or decompression of a gas, and it involves no heat transfer or entropy production. The changing temperature leads to the emission of heat pulses that travel towards the quantum point contact, where the electric current and the heat current are measured. The heat pulses are charge-neutral and generate no average electric current, only noise. On the other hand, a finite heat current passes through the quantum point contact as shown in Fig.~\ref{fig:fig1}(b). The light field generates both a left-going and a right-going heat pulse, carrying opposite energies, but our focus is on the heat pulse that is emitted towards the quantum point contact. That is similar to how a lorentzian voltage pulse generates both a right-going leviton and a left-going antilevition, carrying opposite charge~\cite{levitov1996,keeling2006,thomas2015}. We illustrate our proposal using tight-binding calculations that include the influence of the time-dependent light field~\cite{schonhammer2007,Thomas:2014,thomas2015,gaury2014}.

The rest of the article is organized as follows. In Sec.~\ref{sec:setup}, we introduce the microscopic description of our setup including the light field. In Sec.~\ref{sec:time_dep_T}, we show that the light field can be used to control the electronic temperature in a time-dependent manner, leading to the emission of heat pulses. In Sec.~\ref{sec:counting}, we describe our theoretical framework for calculating the time-dependent currents and their fluctuations using methods from full counting statistics. In Sec.~\ref{sec:constT}, we evaluate the transport properties with a constant light-induced temperature difference between the electrodes, and we show that our results are consistent with scattering theory. In Sec.~\ref{sec:heatpulses}, we consider time-dependent temperatures, which lead to the emission of charge-neutral heat pulses as we show. In Sec.~\ref{sec:conclu}, we present the conclusions of our work together with an outlook on possible avenues for future developments. Several technical details are deferred to Appendices~\ref{app:continuous},~ \ref{app:Riccati},~and~\ref{app:noise}.

\begin{figure*}
	\centering
	\includegraphics[width = 0.98\textwidth]{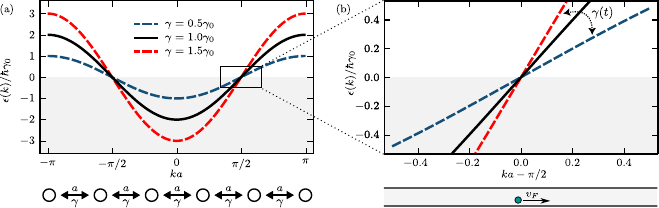}
	\caption{Changing the Fermi velocity using a light field. (a) The standard cosine band structure of a tight-binding chain for different values of the tunnel coupling $\gamma$. The tunnel coupling together with the distance between the atoms in the chain determines the Fermi velocity $v_F=2\gamma a$. We consider a half-filled chain, such that states up until $\epsilon(k)=0$ are occupied. (b)~Close to the Fermi energy, the dispersion relation is linear, and a continuum description of the tight-binding chain applies. A fast external light field allows us to change the tunnel coupling $\gamma(t)$ in parts of the system and thereby the Fermi velocity.}
	\label{fig:fig2}
\end{figure*}

\section{Generation of heat pulses}
\label{sec:setup}

We consider the setup in Fig.~\ref{fig:fig1}(a), showing a left and a right electrode connected by a quantum point contact. To generate heat pulses, we shine light on the left electrode. The full Hamiltonian of the setup reads
\begin{equation}\label{eq:h_tight-binding}
	\hat H(t) = \hat H_L(t) +  \hat H_R+\hat H_\mathrm{qpc},
\end{equation}
where both electrodes consist of a chain of $n$ atoms, which are described by the tight-binding Hamiltonians
\begin{equation}\label{eq:H_L}
	\hat H_L(t) = -\hbar\gamma\sum_{l=1}^{n-1} (e^{i\phi_l(t)}\hat c_l^\dagger \hat c_{l+1}^{\phantom{\dagger}} + e^{-i\phi_l(t)}\hat c_{l+1}^\dagger \hat c_l^{\phantom{\dagger}})
\end{equation}
and
\begin{equation}\label{eq:H_R}
	\hat H_R = -\hbar\gamma\sum_{l=n+1}^{2n-1} (\hat c_l^\dagger \hat c_{l+1}^{\phantom{\dagger}} + \hat c_{l+1}^\dagger \hat c_l^{\phantom{\dagger}}).
\end{equation}
Here, the electronic annihilation operator for atom number $l$ is denoted by $\hat c_l$, and $\gamma$ is the tunnel coupling between neighboring atoms separated by the distance~$a$. The total length of the system is then $L=2n a$. The additional phase $\phi_l(t)$ in Eq.~\eqref{eq:H_L} describes the influence of the light field as we discuss below. Without the light field, electrons close to the Fermi level propagate at the velocity $v_F = 2 \gamma a$. The electrodes are coupled by a quantum point contact described by the Hamiltonian
\begin{equation}\label{eq:H_QPC}
	\hat H_\mathrm{qpc} = -\hbar\gamma_{\mathrm{qpc}}(\hat c_{n}^\dagger \hat c_{n+1}^{\phantom{\dagger}} + \hat c_{n+1}^\dagger \hat c_{n}^{\phantom{\dagger}}),
\end{equation}
where the tunnel coupling $\gamma_{\mathrm{qpc}}\leq\gamma$ determines the transmission of the quantum point contact. 

We now consider the external light field, which is described as a monochromatic plane wave that is polarized along the chain of atoms. The component of the vector potential along the chain then reads
\begin{equation}
A(x,t) = A_0(t)\sin(\Omega t-2\pi x/\lambda),
\end{equation}
where $\Omega$ is the frequency of the light, and $\lambda = 2\pi c/\Omega$ is the corresponding wavelength with $c$ being the speed of light. We also allow for a slowly varying amplitude $A_0(t)$ of the light field. In practice, the wavelength can be on the order of hundreds of nanometers, which is much larger than the typical lattice spacing, $\lambda\gg a$~\cite{merboldt2025}. We can therefore treat the field as spatially uniform and take $A(x,t) = A_0(t)\sin(\Omega t)$ in the irradiated region.

The light field is now included through a Peierls substitution, which gives rise to the phase factor~\cite{peierls1933}
\begin{equation}
\label{eq:peierls}
	\phi_l(t)=\frac{e}{\hbar}\int_{la}^{la+a}dx A(x,t)=\alpha(t)\sin(\Omega t),
\end{equation}
where the dimensionless driving amplitude reads
\begin{equation}
\alpha(t) = eaA_0(t)/\hbar.
\end{equation}
The light field is applied to parts of the left electrode, where the phase factor is then given by Eq.~\eqref{eq:peierls}. Outside this region, the phase factor is zero, $\phi_l(t)=0$.

The Hamiltonian in Eq.~\eqref{eq:h_tight-binding} forms the basis for our calculations in the following. However, before calculating the charge and heat transport in response to the light field, we first analyze the expected impact of the drive.

\section{Time-dependent temperature}
\label{sec:time_dep_T}

In the following, we assume that the frequency of the light is much larger than the tunnel coupling, $\Omega \gg \gamma$. By contrast, the modulations of the driving amplitude are slow. In this limit, electrons that tunnel between neighboring atoms experience a period-averaged light field, such that the effective tunnel coupling becomes
\begin{equation}
	\gamma(t)\simeq\frac{\gamma}{\mathcal T}\int_0^{\mathcal T} ds \exp{(i\alpha(t)\sin(\Omega s))} =  J_0 (\alpha(t))\gamma,
\end{equation}
where $\mathcal T=2\pi/\Omega$ is the period of the drive, and
$J_0$ is the zeroth Bessel function. Here, we have assumed that the driving amplitude, $\alpha(t)$, is roughly constant during a period of the drive. Thus, by modifying the driving amplitude, the effective tunnel coupling can be controlled in a time-dependent manner. For instance, by taking $\alpha(t)\simeq 2.4$, so that the Bessel function vanishes, the effective tunnel coupling can even be reduced to zero, which is known as coherent destruction of tunneling~\cite{grossmann1991,camalet2003}.

Figure~\ref{fig:fig2}(a) illustrates how the band structure of the irradiated region is modified by the changing tunnel coupling. In particular, close to the Fermi energy, the modified tunnel coupling tilts the linear dispersion relation, as seen in Fig.~\ref{fig:fig2}(b). Right-moving electrons close to the Fermi level can then be described by the Hamiltonian
\begin{equation}
\label{eq:LutField}
    \hat H_0(t)= J_0 (\alpha(t))\hat H_0,
\end{equation}
where $\hat H_0=\hbar v_F \hat{q}$ is the Hamiltonian without the light, and we have defined $\hat{q}=\hat{k}-\pi/(2a)$, see~App.~\ref{app:continuous}.

The Bessel function in Eq.~\eqref{eq:LutField} is similar to a Luttinger field, which couples to the total energy and thereby changes the temperature in the irradiated region~\cite{eich2014,tatara2015,biele2015,eich2016,lozej2018,covito2018,karaslimane2020,portugal2024,lopez2024,acciai2025,luttinger1964}. To see this, we consider the density matrix of the irradiated region, assuming first that it is uncoupled from the rest of the system. Initially, the electrons are then described by the thermal density matrix $\hat\rho(0)=\exp(-\hat H_0/k_BT_0)/Z$, where $T_0$ is the temperature of the environment, and $Z$ is the partition function. Now, time-evolving the density matrix with the Hamiltonian in Eq.~\eqref{eq:LutField}, we find $\hat\rho(t)=\hat U(t)\hat\rho(0)\hat U^\dagger(t)=\hat\rho(0)$, where $\hat U(t)$ is the time evolution operator, and we have used the fact that the Hamiltonian commutes with itself at different times. Thus, the density matrix does not change with time. Still, the Hamiltonian changes, and we express the density matrix as $\hat\rho(t)=\exp(-\hat H_0(t)/k_BT(t))/Z$, where
\begin{equation}
\label{eq:inducedT}
T(t) = J_0(\alpha(t)) T_0
\end{equation}
is now the electronic temperature. Thus, by shining light on the electrode, the electronic temperature changes because of the time-dependent tunneling rate.

The temperature changes because we perform work on the electrode. This situation is similar to an adiabatic compression of a gas in a cylinder, whose volume is changed by a piston, which performs work against the pressure. If the compression is fast, there is no time for heat to escape, and the process is reversible and adiabatic in the thermodynamic sense that no entropy is produced. In our case, we can similarly change the electronic temperature using the external light field. Once the irradiated region is coupled to the rest of the system, the changing temperature will generate heat pulses.

\section{Charge and heat transport}
\label{sec:counting}
We now investigate the transport of charge and heat through the quantum point contact in response to the applied light field. To this end, we consider the charge
\begin{equation}
    \hat Q_R = e\sum_{l=n+1}^{2n} \hat c_l^\dagger \hat c_l^{\phantom{\dagger}}
\end{equation}
and the energy in the right electrode given by $\hat H_R$. The corresponding current operators, defined as
\begin{equation}
\begin{split}
\hat I_q =&(i/\hbar)[\hat H, \hat Q_R],\\
\hat I_h =&(i/\hbar)[\hat H, \hat H_R],
\end{split}
\end{equation}
where $\hat H$ is the Hamiltonian without the light field, read
\begin{equation}
\begin{split}
    \hat I_q =& ie\gamma_\mathrm{qpc} (\hat c_n^\dagger \hat c_{n+1}^{\phantom{\dagger}}-\hat c_{n+1}^\dagger \hat c_{n}^{\phantom{\dagger}}),\\
 \hat I_h =& i\hbar \gamma \gamma_\mathrm{qpc} (\hat c_{n}^\dagger \hat c_{n+2}^{\phantom{\dagger}}-\hat c_{n+2}^\dagger \hat c_{n}^{\phantom{\dagger}}).
\end{split}
\end{equation}
In the Heisenberg picture, they are related as $\hat Q_R(t)-\hat Q_R(0) = \int_0^t ds \hat I_q(s)$ and $\hat H_R(t)-\hat H_R(0) = \int_0^t ds \hat I_h(s)$.

In addition to the average currents, we wish to investigate their fluctuations using full counting statistics. Thus, we define the modified Hamiltonian~\cite{levitov1996, esposito2009,hofer2016}
\begin{equation}
\hat H^\nu_\lambda(t)=\hat H(t) + \hbar\lambda \hat I_\nu/2,
\end{equation}
where the counting field, $\lambda$, couples to either the electric current or the heat current, $\nu=q,h$. The equation of motion for the corresponding density matrix reads
\begin{equation}
\label{eq:eomHlam}
\begin{split}
    i\hbar\partial_t \hat\rho^\nu_\lambda(t) =& \hat H^\nu_\lambda(t)\hat\rho^\nu_\lambda(t) - \hat\rho^\nu_\lambda(t) \hat H^\nu_{-\lambda}(t)\\
    =& [\hat H(t),\hat\rho^\nu_\lambda(t)] +\hbar\lambda \{\hat I_\nu,\hat\rho^\nu_\lambda(t)\}/2.
\end{split}
\end{equation}
The density matrix is not normalized, and its trace yields the moment generating function,
\begin{equation}
	M^\nu_\lambda(t) = \mathrm{tr} \{\hat\rho^\nu_\lambda(t)\},
\end{equation}
of the charge or the energy, whose moments are given by the derivatives, $\mu^\nu_m(t) = (-i)^m\partial_\lambda^m M^\nu_\lambda(t)\vert_{\lambda\rightarrow 0}$. We also define the cumulant generating function
\begin{equation}
	C_\lambda^\nu(t) = \ln M^\nu_\lambda(t)
\end{equation}
together with the corresponding cumulants
\begin{equation}
\kappa_m^\nu(t) = (-i)^m\partial_\lambda^m C_\lambda^\nu(t)\vert_{\lambda\rightarrow 0}.
\end{equation}

To find the cumulants, we use that both the Hamiltonian and the observables of interest are quadratic in the fermionic operators. We can then express them as
\begin{equation}
\hat H(t) = \sum_{kl} \mathbb H_{kl}(t) \hat c_k^\dagger \hat c_l^{\phantom{\dagger}}
\end{equation}
and
\begin{equation}
\hat I_\nu = \sum_{kl}\mathbb J_{kl}^\nu \hat c_k^\dagger \hat c_l^{\phantom{\dagger}},
\end{equation}
where both $\mathbb H$ and $\mathbb  J^\nu$ are $2n$-by-$2n$ matrices.

As we show in App.~\ref{app:Riccati}, the equation of motion for the cumulant generating function then becomes
\begin{equation}
\label{eq:cgf}
        i\partial_t C_\lambda^\nu                = \lambda\mathrm{tr}\{\mathbb J^\nu (\mathbb I - \mathbb A_\lambda)\},
\end{equation}
where $\mathbb I$ is the identity matrix. Here, the covariance matrix, $\mathbb A_\lambda$,  evolves according to the Riccati equation~\cite{portugal2023,Kansanen:2025}
\begin{equation}
\label{eq:riccati}
		i \hbar\partial_ t\mathbb A_\lambda = \mathbb H \mathbb A_\lambda - \mathbb A_\lambda \mathbb H +\hbar\lambda(\mathbb J^\nu-\mathbb A_\lambda \mathbb J^\nu \mathbb A_\lambda )/2,
\end{equation}
given the initial condition (at $t=0$)
\begin{equation}
    \mathbb A_\lambda(0)  = \tanh(\mathbb H(0)/2k_BT_0),
\end{equation}
where $[\mathbb A_\lambda(0)]_{kl}  = \langle [\hat c_k^{\phantom{\dagger}},\hat c_l^\dagger]\rangle$ are thermal averages at the initial temperature, and the chemical potential is zero, $\mu=0$, since we work at half filling. 

\begin{figure}
	\centering
\includegraphics[width=0.95\columnwidth]{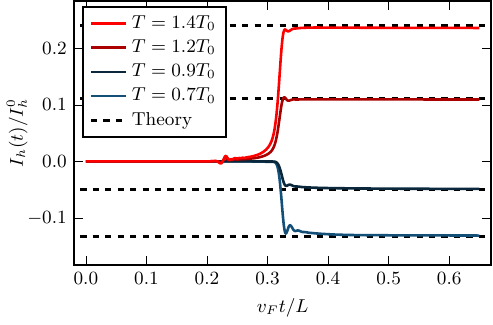}
	\caption{Heat current in response to a constant light-induced temperature difference. The dashed lines are the predictions based on Eq.~\eqref{eq:I_h}, where, for each curve, the driving amplitude $\alpha_0$ has been chosen to give a specific value of $T$. The other parameters are $2n = 1000$, $\Omega = 2.5\gamma$, and $k_BT_0 = 0.025\hbar \gamma /n$, and we have defined $I_h^0 =k_B^2T_0^2/\hbar$ and chosen $\gamma_{\mathrm{qpc}}$, so that $D = 1/2$. The first 225 atoms of the left electrode are irradiated by the light field.}
	\label{fig:fig3}
\end{figure}

\section{Constant temperature}
\label{sec:constT}

\begin{figure*}
	\centering	\includegraphics[width=\textwidth]{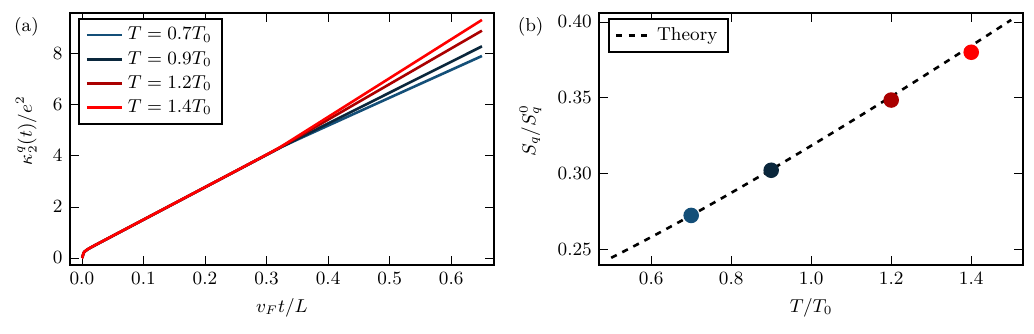}
	\caption{Electric noise in response to a constant light-induced temperature difference. (a) The time-dependent second cumulant of the transferred charge corresponding to the results in Fig.~\ref{fig:fig3}. The long-time behavior of the second cumulant determines the zero-frequency noise of the current. (b) Zero-frequency noise of the current for different light-induced temperatures. The dashed line is the prediction based on Eq.~\eqref{eq:S_I}. The parameters are the same as in Fig.~\ref{fig:fig3}, and we have defined $S_q^0 = e^2 k_BT_0/\hbar$.}
	\label{fig:fig4}
\end{figure*}

We are now ready to illustrate how heat currents are generated in response to the applied light field. We first consider the situation where the light field has a constant amplitude. We then calculate the corresponding charge and heat currents by solving Eqs.~\eqref{eq:cgf} and \eqref{eq:riccati} for the time-dependent Hamiltonian in Eq.~\eqref{eq:h_tight-binding}. We find that the electric current vanishes (not shown), while results for the time-dependent heat current are presented in Fig.~\ref{fig:fig3}.

The figure shows the heat current for different temperatures induced by the light field. Temperatures below the base temperature, $T\leq T_0$, are readily reached according to Eq.~\eqref{eq:inducedT}, since $J_0(\alpha) \leq 1$. On the other hand, to increase the temperature above $T_0$, we prepare the system with the light field already on, assuming that the irradiated part of the electrode has equilibrated at the temperature $T_0$. We then turn off the light field, so that the temperature increases to $T=  T_0/J_0(\alpha) \geq T_0$.

Once the temperature difference is induced by the light field, the heat excitations travel towards the quantum point contact, where they generate a constant heat current as seen in Fig.~\ref{fig:fig3}. We can compare our results with the well-known expression for the heat current through a quantum point contact with transmission~$D$~\cite{acciai2025},
\begin{equation}
I_h= D\frac{\pi^2 k_B^2}{6h}(T^2 - T_0^2),
\label{eq:I_h}
\end{equation}
where $T$ and $T_0$ are the temperatures in the leads. The transmission at low energies can be approximated as~\cite{Thomas:2014}
\begin{equation}
D=\frac{4 \gamma ^2 \gamma_\mathrm{qpc}^2}{(\gamma ^2+\gamma_\mathrm{qpc}^2 )^2},
\end{equation}
while for the temperature in the irradiated region, we take $T=J_0(\alpha)T_0$ or $T=T_0/J_0(\alpha)$ for temperatures below or above the base temperature $T_0$. With these identifications, we find good agreement between our tight-binding results and Eq.~\eqref{eq:I_h}, which is shown with a dashed line. This agreement supports the interpretation that the light field changes the electronic temperature.

To further corroborate this interpretation, we consider the fluctuations of the currents.
Although the average electric current vanishes, the temperature difference still generates fluctuations, also known as $\Delta T$-noise~\cite{lumbroso2018}. With a constant temperature bias, the zero-frequency noise can be extracted from the second cumulant of the transferred charge or energy at long times, see App.~\ref{app:noise},
\begin{equation}
    S_{\nu} = \lim_{t\rightarrow \infty}\frac{d}{dt}\kappa_2^{\nu}(t), \,\,\nu=q,h.
\end{equation}

In Fig.~\ref{fig:fig4}(a), we show the time-evolution of the second cumulant of the charge, which becomes linear in time after an initial transient behavior. We can then extract the zero-frequency noise of the electric current for different light-induced temperatures as shown in Fig.~\ref{fig:fig4}(b). Again, we can compare with predictions based on scattering theory, for which the zero-frequency noise reads~\cite{blanter2000}
\begin{equation}
\label{eq:S_I}
	S_q = \frac{2 e^2}{h} \int d\epsilon \sum_{kl}
	\mathcal{T}_{k l}  f_k(\epsilon)(1 - f_l(\epsilon)),
\end{equation}
where the indices, $k,l=L,R$, run over the left and right leads with Fermi functions $f_k(\epsilon)$, and we have defined $\mathcal{T}_{kk} =  D^2$
and $\mathcal{T}_{k\neq l} = D(1-D)$. The prediction based on Eq.~\eqref{eq:S_I} is shown with a dashed line in Fig.~\ref{fig:fig4}(b), and it agrees well with our tight-binding results.

\begin{figure*}
	\centering	\includegraphics[width=\textwidth]{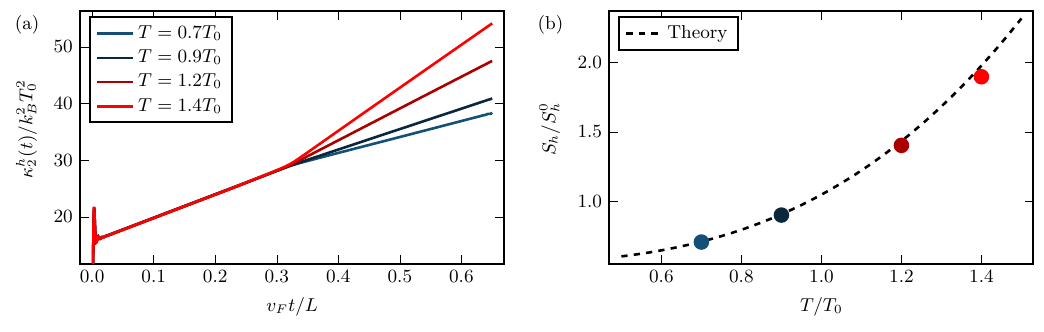}
	\caption{Heat noise in response to a constant light-induced temperature difference. (a) The time-dependent second cumulant of the transferred heat corresponding to the results in Fig.~\ref{fig:fig3}. The long-time behavior of the second cumulant determines the zero-frequency heat noise. (b) Zero-frequency heat noise for different light-induced temperatures. The dashed line is the prediction based on Eq.~\eqref{eq:S_J}. The parameters are the same as in Fig.~\ref{fig:fig3}, and we have defined $S_h^0 = k_B^3 T_0^3/\hbar$.}
	\label{fig:fig5}
\end{figure*}

Similar results for the second cumulant of the heat and the zero-frequency heat noise are shown in Fig.~\ref{fig:fig5}. Based on scattering theory, the zero-frequency heat noise reads
\begin{equation}\label{eq:S_J}
	S_h = \frac{2}{h} \int d\epsilon \epsilon^2 \sum_{kl}
	\mathcal{T}_{k l} \, f_k(\epsilon) (1 - f_l(\epsilon)),
\end{equation}
and also in this case, we find good agreement between scattering theory and our tight-binding calculations. In combination, these results show that the light field indeed changes the temperature of the irradiated region in a predictable and controllable manner.

\section{Temperature pulses}
\label{sec:heatpulses}

We now consider time-dependent pulses, and we thus take a light field, whose amplitude changes in time. As an example, we consider a Gaussian amplitude reading
\begin{equation}
\label{eq:gaussian}
    \alpha(t) =  \frac{\alpha_0}{\sqrt{2\pi}\sigma }  \exp(-t^ 2/ 2 \sigma^2),
\end{equation}
where the width of the pulse is denoted by $\sigma$, and the prefactor $\alpha_0$ controls the amplitude of the pulse.

For a pulse that cools the electrode, the system starts in equilibrium without the light field. The Gaussian pulse is then applied, and the time-dependent temperature becomes $T(t) = J_0(\alpha(t)) T_0 < T_0$. By contrast, for a pulse that heats the electrode, the system is prepared with the light field on, and it is thermalized at the temperature~$T_0$. The Gaussian amplitude is then subtracted, so that the temperature increases to $T(t) = T_0/J_0(\alpha(t)) > T_0$.

\begin{figure*}
	\centering
\includegraphics[width=\textwidth]{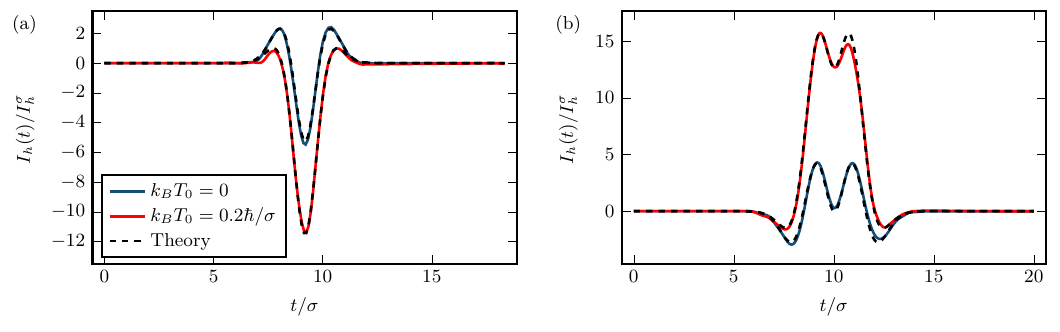}
	\caption{Time-dependent heat current generated by a light pulse. (a) The shape of the light pulse is given by Eq.~\eqref{eq:gaussian}, and we consider two different base temperatures. The dashed lines are the expected heat currents based on Eq.~\eqref{eq:heat_current}. (b) A pulse that increases the temperature. The parameters are $2n = 600$, $\Omega = 35\gamma$, $\alpha_0 = 3$, and we have defined $I^\sigma_h = 10^{-3}\hbar/\sigma^2$. We have chosen $\gamma_{\mathrm{qpc}}$, so that $D = 1/2$, and the first 105 atoms of the left electrode are irradiated by the light field.}
	\label{fig:fig6}
\end{figure*}

Figure~\ref{fig:fig6} shows the heat current generated by a Gaussian light pulse that either cools or heats the electrode. The heat pulse generated by the light field travels at the Fermi velocity towards the quantum point contact, where the heat current is measured. In Fig.~\ref{fig:fig6}(a), the light field reduces the temperature below $T_0$, and the heat current from the left to the right electrode is negative. By contrast, in Fig.~\ref{fig:fig6}(b), the temperature increases above $T_0$, and the heat current is positive. In both panels, we show results for the base temperature being zero or finite.

We can compare our results with predictions based on Floquet scattering theory~\cite{portugal2024}. The heat pulses are charge-neutral, such that no electric current is produced. The heat current, on the other hand, should read
\begin{equation}
\label{eq:heat_current}
	I_h(t) = D\frac{\pi^2 k_B^2}{6h}(T^2(t) - T_0^2) - D\frac{\hbar}{24\pi}(S\tau)(t),
\end{equation}
where $\tau(t) = \int_0^t ds T(s)/T_0=\int_0^t ds J_0(\alpha(s))$ can be considered as a dilated time, and $(S\tau)(t) = \dddot{\tau}/\dot{\tau} - 3(\ddot{\tau}/\dot{\tau})^2/2$ is the Schwarzian derivative of $\tau(t)$~\cite{gawedzki2018, bermond2024}. The first term is the static result from Eq.~\eqref{eq:I_h} with the time-dependent temperature inserted. This term dominates at high temperatures, but vanishes as the base temperature approaches zero. The second term, by contrast, is a high-frequency quantum correction that is present even at zero temperature. It can also be derived using conformal field theory~\cite{gawedzki2018}, and it ensures that the fluctuation-dissipation theorem for heat currents is fulfilled~\cite{portugal2024}. Without it, the thermal conductance would be frequency-independent, and it would appear as if the fluctuation-dissipation theorem is violated~\cite{Averin:2010}.

In Fig.~\ref{fig:fig6}, we show the predictions based on Eq.~(\ref{eq:heat_current}) with dashed lines, and we find good agreement with our tight-binding calculations. The results at zero temperature can be fully associated with the Schwarzian derivative in Eq.~(\ref{eq:heat_current}), while the results at a finite base temperature also involve the static term with the time-dependent temperature inserted. Now, we can also return to Fig.~\ref{fig:fig1}(b), which is similar to Fig.~\ref{fig:fig6}(b), however, in Fig.~\ref{fig:fig1}(b), the results were obtained at the elevated base temperature of $k_BT_0 =  1.0\hbar/\sigma$ for a chain of length~$2n=300$. Based on these calculations, we can conclude that the time-dependent light field indeed changes the electronic temperature as expected.

Finally, we comment on realistic parameters. Typical Fermi velocities are around $v_F\simeq 10^5$~ms$^{-1}$ with interatomic spacings of about $a\simeq0.2$~nm~\cite{joyce2013,hao2026}. The tunnel coupling is then $\gamma \simeq 250$~THz. For the frequency of the light, one may use $\Omega \simeq 1000$~THz, corresponding to a wavelength of about $\lambda=300$~nm (ultraviolet light) to ensure that~$\Omega>\gamma$. The modulations of the amplitude, on the other hand, can occur at much lower frequencies, on the order of 1--10~GHz, corresponding to an electronic temperature between 50~mK and 500~mK, which is readily reached in realistic solid-state experiments.

\section{Conclusions \& Outlook}
\label{sec:conclu}

We have proposed to generate heat pulses in mesoscopic conductors using light fields. The central idea is that a high-frequency light field renormalizes the tunnel coupling between neighboring atoms in an electrode and thereby changes the Fermi velocity. Electrons close to the Fermi level can then be described by a Hamiltonian with a time-dependent Luttinger field that changes the electronic temperature in the irradiated region. The process is coherent and involves no entropy production. Once the irradiated electrode is coupled to a conductor, the time-dependent temperature generates heat pulses that produce a heat current but no electric current.

We have validated this picture using tight-binding calculations combined with methods from full counting statistics. For a constant light-induced temperature difference, our results for both the heat current and the zero-frequency current fluctuations are in excellent agreement with predictions from scattering theory, confirming that the light field modifies the electronic temperature in an expected and controllable manner. For time-dependent light amplitudes, we have shown that the resulting heat current is well captured by Floquet scattering theory, which corroborates the idea of our proposal.

Our work establishes a concrete and experimentally accessible route towards on-demand caloritronics, in which energy rather than charge serves as the carrier of quantum information. The proposed scheme requires only a pulsed light source and a standard mesoscopic conductor, with realistic parameters pointing to ultraviolet light and electronic temperatures in the subkelvin regime.

Several directions for future research present themselves naturally. Our work is based on a non-interacting theory, and it is an open question how Coulomb interactions would modify the shape and coherence of the emitted heat pulses, in particular given that the heat pulses are charge-neutral. Extensions to topological edge states and fractional quantum Hall systems, where the thermal conductance takes half-integer values, would be an obvious testing ground for the interplay between bulk topology and the coherence of thermally driven edge excitations.
Finally, the Schwarzian derivative appearing in the heat current hints at deeper connections between dynamic heat transport and conformal field theory, which we leave as an interesting topic for further investigation.

\acknowledgments{
We thank M.~Saha for useful discussions and acknowledge the financial support of the Research Council of Finland through the Finnish Quantum Flagship (Project Nos.~358877 and 359240). This article is based upon work from COST Action ``Many-body Open Quantum Systems'' (QOpen) CA24109, supported by COST (European Cooperation in Science and Technology). R.~T.~acknowledges the financial support of the Jane and Aatos Erkko Foundation (Project EffQSim).
}

\appendix

\section{Continuum description}
\label{app:continuous}

In this section, we establish the low-energy continuum description of a tight-binding chain with the dispersion relation $\epsilon(k) = -2 \hbar \gamma \cos(k a)$. Here, the lattice spacing is denoted by $a$, and $\gamma$ is the tunnel coupling. We assume that the tight-binding chain is half-filled, so that the chemical potential is zero. The Fermi momentum is then $k_F = \pm \pi/2a$, since $\epsilon(k_F) = -2\hbar \gamma \cos(k_F a) = 0$.

We are interested in excitations close to the Fermi level. We therefore define $q =k\mp k_F $ with $|q|\ll \pi/a$. We then introduce separate fields, $\hat \psi_\alpha$, for right- and left-movers,
\begin{equation}
	\hat c_l \simeq\sqrt{a}[e^{ik_F x_l} \hat \psi_R (x_l) + e^{-ik_F x_l} \hat \psi_L (x_l)],
\end{equation}
where $x_l$ is the position of site $l$ in the chain. These operators obey the usual anticommutation relations
\begin{equation}
\{\hat \psi_\alpha(x), \hat \psi_\beta^{\dagger}(x')\} = \delta_{\alpha\beta} \delta(x - x'),\quad \alpha,\beta=L,R.
\end{equation}
The tight-binding model Hamiltonian can now be written
\begin{equation}
	\hat H = -i \hbar v_F\sum_{l=1}^{n-1}  \hat\Psi^\dagger(x_l) \sigma_z (\hat\Psi(x_{l+1}) - \hat\Psi(x_l)),
\end{equation}
where $\sigma_z$ is the third Pauli matrix, and we have defined 
\begin{equation}
\hat\Psi(x) = \left(
\begin{matrix}
\hat \psi_{R}(x)\\
\hat \psi_L(x)
\end{matrix}
\right).
\end{equation}
Considering the lattice spacing, $a$, to be short compared to other relevant length scales, we get
\begin{equation}\label{eq:h_ballistic}
	\hat H = -i \hbar v_F \int_{0}^L dx \hat\Psi^\dagger(x)\, \sigma_z \partial_x \hat\Psi(x),
\end{equation}
where $L$ is the length of the tight-binding chain. In momentum space, the Hamiltonian then becomes
\begin{equation}\label{eq:h_ballistic_kspace}
	\hat H = \pm\hbar v_F  \sum_q q \hat c_q^\dagger \hat c_q =\pm\hbar v_F \hat q^{\phantom{\dagger}}
\end{equation}
with different signs for left- and right-movers.

\section{Full counting statistics}
\label{app:Riccati}

To evaluate the full counting statistics of transferred charge or heat, we write Eq.~\eqref{eq:eomHlam} on the form
\begin{equation}
		i\hbar\partial_t \hat\rho^\nu_\lambda  =\sum_{kl} \mathbb H_{kl} [\hat c_k^\dagger \hat c_l^{\phantom{\dagger}}, \hat\rho^\nu_\lambda] + \hbar\lambda\mathbb J^{\nu}_{kl} \{\hat\rho^\nu_\lambda, \hat c_k^\dagger \hat c_l^{\phantom{\dagger}}\}/2.
        \label{eq:EOMrho}
\end{equation}
Using third quantization, we define the operators~\cite{mcdonald2023,Kim:2023}
\begin{equation}
	\begin{split}
		\mathbf c_{1l}^{\phantom{\dagger}}|\hat\rho\rangle          = \frac{1}{\sqrt{2}}|\{\hat c_l^{\phantom{\dagger}},\hat\rho \hat P\}\rangle,&\quad  \mathbf c_{1l}^\dagger|\hat\rho\rangle  = \frac{1}{\sqrt{2}}|[\hat c_l^\dagger,\hat\rho\hat P]\rangle,         \\
		\mathbf c_{2l}^{\phantom{\dagger}}|\hat\rho\rangle          = \frac{1}{\sqrt{2}}|[\hat c_l,\hat\rho\hat P]\rangle,& \quad
		\mathbf c_{2l}^\dagger|\hat\rho\rangle  = \frac{1}{\sqrt{2}}|\{\hat c_l^\dagger,\hat\rho\hat P\}\rangle,
	\end{split}
\end{equation}
having introduced the parity operator,
\begin{equation}
\hat P = \exp(i \pi \sum_l \hat c_l^\dagger \hat c_l^{\phantom{\dagger}}),
\end{equation}
and a bracket notation for density matrices. These operators satisfy the anticommutation relations
\begin{equation}
	\{\mathbf c_{ak}^{\phantom{\dagger}}, \mathbf c_{bl}^\dagger\} = \delta_{ab}^{\phantom{\dagger}} \delta_{kl}^{\phantom{\dagger}} \mathbf{1},\quad \{\mathbf c_{ak}^{\phantom{\dagger}}, \mathbf c_{bl}^{\phantom{\dagger}}\} = \mathbf 0.
\end{equation}
In addition, the inner product is defined as $\langle \hat A | \hat B \rangle = \mathrm{tr}\{\hat A^\dagger \hat B\}$. Finally, we 
will need the identity
\begin{equation}
| 0_1 1_2\rangle = \prod_l \mathbf c_{2l}^\dagger |0\rangle,
\end{equation}
where $\langle 0_1 1_2 | \hat \rho\rangle = \mathrm{tr}\{\hat\rho\}$, and $\mathbf c_{1l}^{\phantom{\dagger}}|0\rangle = \mathbf c_{2l}^{\phantom{\dagger}}|0\rangle = 0$ defines the vacuum state $|0\rangle$. We can now express Eq.~\eqref{eq:EOMrho} as
\begin{equation}\label{eq:liouvillian}
	\begin{split}
		i\hbar\partial_t |\hat\rho^\nu_\lambda\rangle =  &\sum_{kl} \mathbb H_{kl} (\mathbf c_{1k}^\dagger \mathbf c_{1l}^{\phantom{\dagger}} - \mathbf c_{2l}^{\phantom{\dagger}} \mathbf c_{2k}^\dagger)   |\hat\rho^\nu_\lambda\rangle                          \\
		             + \frac{\hbar\lambda}{2}&\sum_{kl}\mathbb J_{kl}^\nu (\mathbf c_{1k}^\dagger \mathbf c_{2l}^{\phantom{\dagger}} - \mathbf c_{1l}^{\phantom{\dagger}} \mathbf c_{2k}^\dagger + \delta_{kl}^{\phantom{\dagger}}\mathbf{1})|\hat\rho^\nu_\lambda\rangle.
	\end{split}
\end{equation}
For the cumulant generating function, we then find
\begin{equation}
i\partial_t C_\lambda^\nu=	\frac{\langle 0_1 1_2 | i\partial_t \hat\rho^\nu_\lambda\rangle}{\langle 0_1 1_2 |\hat\rho^\nu_\lambda \rangle} = \lambda\mathrm{tr}\{\mathbb J^\nu (\mathbb I - \mathbb A_\lambda)\},
\end{equation}
where the antisymmetric covariance matrix is defined as
\begin{equation}
	[\mathbb A_\lambda(t)]_{kl} = \frac{\langle [\hat c_k^{\phantom{\dagger}},\hat c_l^\dagger]\rangle(t)}{M^\nu_\lambda(t)} = \frac{\langle 0_1 1_2 |  \mathbf c_{1k}^{\phantom{\dagger}}\mathbf c_{2l}^\dagger | \hat\rho^\nu_\lambda(t)\rangle}{M^\nu_\lambda(t)}.
\end{equation}
The equation of motion for the covariance matrix reads
\begin{equation}
\label{eq:covmatrix}
	\begin{split}
		i \hbar\partial_ t[\mathbb A_\lambda(t)]_{kl} & =  \langle 0_1 1_2 | \mathbf c_{1k}^{\phantom{\dagger}}  \mathbf c_{2l}^\dagger \left|i \hbar\partial_t \left(\hat\rho^\nu_\lambda(t)/M^\nu_\lambda(t)\right)\right\rangle \\
		& =  [ \mathbb H \mathbb A_\lambda - \mathbb A_\lambda \mathbb H +\hbar\lambda(\mathbb J^\nu-\mathbb A_\lambda \mathbb J^\nu \mathbb A_\lambda )/2]_{kl},
	\end{split}
\end{equation}
where $\hat\rho^\nu_\lambda$ is assumed to be a Gaussian state.

\section{Zero-frequency noise}
\label{app:noise}

In this section, we show how the zero-frequency noise of the electric current and the heat current can be related to the second cumulant of the transferred charge or heat. To this end, we rewrite Eq.~\eqref{eq:eomHlam} as
\begin{equation}
    i\partial_t \tilde\rho^\nu_\lambda(t) = \lambda\{\tilde{I}_\nu(t),\tilde\rho^\nu_\lambda(t)\}/2,
    \label{eq:eom_int}
\end{equation}
having defined
\begin{equation}
\begin{split}
    \tilde\rho^\nu_\lambda(t)& = \hat{U}^\dagger(t,t_0)\hat\rho^\nu_\lambda(t)\hat{U}(t,t_0),\\ \tilde{I}_\nu(t) &= \hat{U}^\dagger(t,t_0)\hat{I}_\nu(t)\hat{U}(t,t_0),
    \end{split}
\end{equation}
where $\hat{U}(t,t_0)$ is the time-evolution operator generated by~$\hat H(t)$. Next, we expand $\tilde\rho^\nu_\lambda(t)$ in $\lambda$ as
\begin{equation}
    \tilde\rho^\nu_\lambda(t) =  \tilde\rho^\nu_{0}+\lambda\,\tilde\rho^\nu_{1}(t) + \lambda^2\,\tilde\rho^\nu_{2}(t) + \ldots
\end{equation}
and collect terms order-by-order in Eq.~\eqref{eq:eom_int} as
\begin{equation}
    \begin{split}
    i\partial_t \tilde\rho^\nu_{1}(t) &= \{\tilde{I}_\nu(t),\tilde\rho^\nu_{0}\}/2, \\
    i\partial_t \tilde\rho^\nu_{2}(t) &= \{\tilde{I}_\nu(t),\tilde\rho^\nu_{1}(t)\}/2,
    \end{split}
\end{equation}
where $\tilde\rho^\nu_{0}$ is the initial condition.

Formally solving these equations, we find
\begin{equation}
\tilde\rho^\nu_{1}(t) = \frac{1}{2i}\int_0^t ds \{\tilde{I}_\nu(s),\tilde\rho^\nu_{0}\},
\end{equation}
and
\begin{equation}
\tilde\rho^\nu_{2}(t) = \frac{1}{(2i)^2}\int_0^t ds_1\int_0^{s_1} ds_2 \{\tilde{I}_\nu(s_1),\{\tilde{I}_\nu(s_2),\tilde\rho^\nu_{0}\}\}.
\end{equation}
The second moment reads $\mu^\nu_2(t)= -2\,\mathrm{tr}\{\tilde\rho^\nu_{2}(t)\}$, and 
using the cyclic property of the trace, we find
\begin{equation}
\begin{split}
    \mu^\nu_2(t) =& \int_0^t ds_1\int_0^{s_1} ds_2\langle\{\tilde{I}_\nu(s_1),\tilde{I}_\nu(s_2)\}\rangle\\
    =& \frac{1}{2}\int_0^t ds_1\int_0^t ds_2 \langle\{\tilde{I}_\nu(s_1),\tilde{I}_\nu(s_2)\}\rangle,
\end{split}
\end{equation}
since the integrand is symmetric under the exchange $s_1\leftrightarrow s_2$. Now, considering the deviations from the mean,
\begin{equation}
\delta\tilde{I}_\nu(t) = \tilde{I}_\nu(t) - \langle\tilde{I}_\nu(t)\rangle,
\end{equation}
and defining by the correlation function
\begin{equation}
\mathcal{C}_\nu(t_1,t_2)=\frac{1}{2}\langle\{\delta\tilde{I}_\nu(t_1),\tilde{I}_\nu(t_2)\}\rangle,
\end{equation}
the second cumulant can be expressed as
\begin{equation}
    \kappa_2^\nu(t) = \int_0^t ds_1 \int_0^t  ds_2\mathcal{C}_\nu(s_1,s_2).
\end{equation}

Assuming that the process eventually becomes stationary, for example, if the heat flow levels out, the correlator will depend only on the time difference, $\mathcal{C}_\nu(t_1,t_2)=\mathcal{C}_\nu(t_1-t_2)$. Next, changing variables to $s = (s_1+s_2)/2$ and $\tau = s_1 - s_2$, and integrating over $s$, we find
\begin{equation}
    \kappa_2^\nu(t) = \int_{-t}^{t} d\tau  (t - |\tau|) \mathcal{C}_\nu(\tau).
\end{equation}
The zero-frequency noise can then be found as
\begin{equation}
 S_\nu(0)\equiv  \int_{-\infty}^{\infty} d\tau  \mathcal{C}_\nu(\tau) = \lim_{t\to\infty} \partial_t \kappa_2^\nu(t).
\end{equation}

%

\end{document}